\documentclass[useAMS,usenatbib,12pt]{mn2e}
\usepackage[latin1]{inputenc}
\usepackage{graphics}
\usepackage{amsfonts}
\usepackage{amsmath}
\usepackage{multicol}
\usepackage{layout}
\usepackage{amssymb}
\usepackage{epsfig}
\usepackage[a4paper]{hyperref}
\def\msun{{\rm\,M_\odot}}

\def\cm{{\rm\,cm}}
\def\sec{{\rm\,s}}

\def\sr{{\rm\,sr}}

\def\cm{{\rm\,cm}}

\def\kpc{{\rm\,kpc}}
\def\pc{{\rm\,pc}}

\def\GeV{{\rm\,GeV}}

\def\sec{{\rm\,s}}
\def\sr{{\rm\,sr}}

\def\B0{B$_{ref,0}$} 
\def\Bz{B$_{ref,z}$}

\newcommand{\beq}{\begin{equation}}
\newcommand{\eeq}{\end{equation}}
\newcommand{\be}{\begin{equation}}
\newcommand{\ee}{\end{equation}}

\long\def\comment#1{}

\def\msun{M_{\odot}}
\newcommand{\ba}{\begin{eqnarray}}
\newcommand{\ea}{\end{eqnarray}}

\def\rsun{{R_\odot}} 

\title[Analytical Subhaloes Population]{Analytical Approach to
                  Subhaloes Population in Dark Matter Haloes}
\author[C. Giocoli, L. Pieri \& G. Tormen]{
Carlo Giocoli$^1$,
Lidia Pieri$^{2,3}$ and
Giuseppe Tormen $^1$ \\
\thanks{Email:
 \href{mailto:carlo.giocoli@unipd.it}{carlo.giocoli@unipd.it},
 \href{mailto:lidia.pieri@oapd.inaf.it}{lidia.pieri@oapd.inaf.it},
 \href{mailto:giuseppe.tormen@unipd.it}{giuseppe.tormen@unipd.it}.
}
\\
$^1$ Dipartimento di Astronomia, Università degli Studi di Padova,
       Vicolo dell'osservatorio 2 I-35122 Padova, Italy \\
$^2$ Istituto Nazionale di Astrofisica - Osservatorio Astronomico di Padova,
       Vicolo dell'osservatorio 2 I-35122 Padova, Italy \\
$^3$ Istituto Nazionale di Fisica Nucleare - Sezione di Padova, Via Marzolo 8 I-35131 Padova, Italy}
\begin{document}
\date{}
\pagerange{\pageref{firstpage}--\pageref{lastpage}} \pubyear{2007}
\maketitle
\label{firstpage}

\begin{abstract}
In the standard model of cosmic structure formation, dark matter 
haloes form by gravitational instability. The process is hierarchical:
smaller systems collapse earlier, and later merge to form larger 
haloes. The galaxy clusters, hosted by the largest dark matter haloes, are 
at the top of this hierarchy representing the largest as well as the last 
structures formed in the universe, while the smaller and first haloes are
those Earth-sized dark subhaloes which have been both predicted by theoretical 
considerations and found in numerical simulations, though it does not exist any 
observational hints of their existence. 
The probability that a halo of mass $m$ at redshift $z$ will be
part of a larger halo of mass $M$ at the present time can be described in the 
frame of the extended Press \& Schecter theory making use of
the progenitor (conditional) mass function.
Using the progenitor mass function we calculate analytically, at
redshift zero, the distribution of subhaloes in mass, formation epoch 
and rarity of the peak of the density field at the formation epoch. 
That is done for a Milky Way-size system, assuming both a spherical
and an ellipsoidal collapse model. Our calculation assumes that 
small progenitors do not lose mass due to dynamical processes after
entering the parent halo, and that they do not interact with other subhaloes. 
For a $\mathrm{\Lambda}$CDM power spectrum we obtain a subhalo mass function 
$\mathrm{d}n/\mathrm{d}m$ proportional to $m^{- \alpha}$ with a 
model-independent $\alpha \sim 2$.
Assuming the dark matter is a weakly interacting massive particle, the inferred 
distributions is used to test the feasibility of an indirect 
detection in the $\gamma$-rays energy band of such a population of subhaloes 
with a GLAST-like satellite.
\end{abstract}
  
\section{Introduction}
The present-day description of the universe includes the presence of 
a large amount of cold dark matter (CDM) whose nature and distribution 
is unknown. This Dark Matter (DM) provides about 26 \% of the energy budget 
of the universe.

The amount and properties of CDM is well constrained by astrophysical 
observations such as the anisotropies in the Cosmic Microwave Background, 
large scale structure and distant type I A supernovae
\citep{spe03,sup06,sdss}. 
On the other hand, two main open questions arise.
The first concerns the particle physics nature of the CDM.
Weakly interacting massive particles (WIMPs) are attractive candidates 
since their relic abundance can fit the observed one \citep{dim:90}.   
Stable neutralinos in supersymmetric extensions of the standard model 
(SUSY) \citep{jun:96,bhs:05} or Kaluza-Klein particles (KKP) in theories 
with a TeV$^{-1}$ size universal extra dimension 
\citep{app01,servanttait} are the most commonly studied particles. 
Since these particles have never been observed, there is a large 
uncertainty on the prediction of their effects which has to be taken 
into account. 
The other open question regards the distribution of DM inside the 
haloes. Numerical N-body simulations \citep{Navarro:97,Moore:04,Navarro:03}, 
whose scale resolution is about $\sim 0.1 \kpc$, allow solely an 
extrapolation of the very inner slope of the DM profile and do not take into
account interactions with the baryons which fall in the DM potential
well or the presence of inner cores \citep{Berezinsky03} or the controverse 
effect of the presence of a black hole at the centre of the halo 
\citep{ullio:01,bertone:05,bertone2:05,Merritt:02}. Experimental data on 
DM distribution in the haloes of galaxies and clusters are not conclusive 
too (see, i.g., the discussion in \citet{fps04}). 
In the hierarchical formation scheme of the CDM scenario, large systems 
are the result of the merging and accretion of smaller haloes
(subhaloes), whose dense central cores would survive the merging 
event and continue to orbit within the parent halo, as shown by 
high resolution N-body simulations \citep{Moore:99,Ghigna:99,Blasi:00}. 
CDM models are characterized by an excess of power on small scales. 
The arising divergence of the linear density constrast at large wavenumbers 
has been proved to be damped by collisional processes and free streaming, 
respectively before and after kinetic decoupling, leading to exponential 
damping of the linear CDM density contrast and to the existence of a typical 
scale (Jeans scale) for the first haloes corresponding to a Jeans mass 
about $10^{-6} \msun$ \citep{hss:01,ghs:04,ghs:05}. 
Numerical simulations have indeed reproduced hierarchical 
clustering in CDM cosmologies with a mass resolution sufficient to
resolve the Jeans mass \citep{dnat05} with particle mass 
$m_p = 1.2 \times 10^{-10}\,M_{\odot}$ and force resolution of 
$\epsilon = 0.01$ pc; however such a high resolution run could be 
evolved only to $z=26$, in a very small spatial patch, and producing
haloes of mass $[10^{-6}, 10^{-4}]\,M_{\odot}$.

Among the simulations evolved on larger scales and to
redshift $z=0$, present milestones are the Millennium Simulation
\citep{millennium} and the Via Lactea Simulation
\citep{dkm07a}. The first is a cosmological N-Body run with over $10$
billion particles in a cubic region $500\,\mathrm{Mpc}/h$ on a side
(particle mass $m_p = 1.23 \times 10^9 M_\odot$; force resolution
$\epsilon = 7$ kpc); the second was done to obtain a simulated Milky
Way with the highest possible mass resolution (particle mass $m_p =
2.09 \times 10^3 M_\odot$; force resolution $\epsilon = 90$ pc).
However a simulation with the mass and force resolution similar to
that of \citep{dnat05}, evolved to redshift zero over a region
containing a mass comparable to that of our Galaxy would require about $10^{20}$ particles and a time resolution of a few years. Such
requirements are way beyond the computational capabilities of
present-day supercomputers: applying Moore's law and starting from
present day state-of-the-art, a run like this could be performed in
roughly $50$ years from now. 

A reasonable alternative is to study the
clustering properties of Milky Way-like systems through an analytical approach.
We use the fact that the probability that a halo of mass $m$
at redshift $z$ will be
part of a larger halo of mass $M$ at the present time is described 
by the progenitor conditional mass function $f(m,z|M,z_{0}=0)$, 
according to the so-called extended Press \& Schechter theory.
Using the progenitor mass function, we can calculate analytically, at
redshift zero, the distribution of subhaloes in mass, formation epoch 
and rarity of the peak of the density field at the formation epoch. 
That is done for a Milky Way-size system, assuming both a spherical
and an ellipsoidal collapse model. 

Numerical simulations described in \citet{detal05} show 
that the distribution of material originating from 
the earliest branches of the merger tree within the present day haloes 
depends on the $\sigma$-peaks of the primordial density fluctuation field 
it belonged to. 
We extend their numerical results by performing an analytical estimate of 
the density peaks distribution as a function of the halo mass traced back 
to the smallest scale haloes, thus avoiding the limitation imposed by numerical 
simulations. In this way we obtain a realistic estimate of the distribution 
and mass function of the whole population of subhaloes.

Such an analytical estimate can provide a powerful tool to take 
into account the effect of early high-density peaks in present day haloes. 

This is particularly important in the framework of dark matter indirect detection, since a high $\sigma$-peak halo translates into a higher concentration and thus a higher value for the density squared which has to be integrated along the line of sight to obtain a prediction for particle fluxes coming from dark matter annihilation. 

Given some model for the hierarchical formation of our Galaxy, and for
the internal structure of subhaloes, DM may be in fact indirectly detected
using annihilation rates predicted from particle physics
\citep{Bergstrom:2000pn,bhs:05} through the observation of high
density point-source or extended regions inside our Galaxy.  If we
restrict ourselves to $\gamma$-ray observations, these can be obtained
using either atmospheric Cerenkov telescopes
\citep{VERITAS,HESS,MAGIC} or satellite-borne detectors like GLAST
\citep{GLAST}.  The detectability of DM substructures with GLAST has
been widely discussed in the literature 
(see, e.g. \citet{PBB07} and references therein). 
The small mass haloes have been found to give the main contribution to 
an unresolved $\gamma$-ray foreground arising from DM annihilation, while their
detection as resolved objects has been proved to be very unlike.
Indeed the unresolved subhalo foreground is prominent above the MW smooth 
foreground far from the Galactic Center, where the overall flux is still too low to be detected.

In this paper we apply the analytical derivation of the subhalo population 
properties, such as the $\sigma$-peak distribution, on the indirect 
detection of $\gamma$-rays.
We thus study the possibility that high $\sigma$-peak material could 
arise the foreground level above the detectability threshold
of a GLAST-like large field of view satellite. 

As in \citet{PBB07}, we use different models for the virial concentration 
of subhaloes.

The paper is organized as follows: in
Sec.~\ref{sec:analyticalapproach}, we review the spherical 
and ellipsoidal collapse model and their properties. 
In Sec.~\ref{sec:USMF} we describe the original analytical derivation
of the density peak distribution as function of the halo mass,
and the subhaloes mass function for a present day halo with mass 
$M=10^{12}\,M_{\odot}/h$.
In Sec.~\ref{sec:flux} we estimate the upper bound for the 
contribution to the $\gamma$-ray flux due to the presence 
of a population of subhaloes inside the Milky Way. In Sec.
\ref{sec:detection}, we study the prospects for detection 
of substructures with a GLAST-like experiments in 
our best case scenario. A discussion of our results can be found 
in Sec.~\ref{sec:conclusions}.

\section{Extended-Press \& Schechter: from Progenitors to Subhaloes}
\label{sec:analyticalapproach}
In the hierarchical picture of galaxy formation, structures up to 
protogalactic scale grow as a consequence of repeated merging events. 
Smaller systems collapse at high redshifts, when the universe is denser, 
and subsequently assemble to form bigger and bigger haloes \citep{lc93}. 
This merging history is often represented by the so called ''merger-trees''. 

Smaller systems accreted onto a larger halo along its merging-history-tree 
and still surviving at a later time are called ''substructures'' or 
''subhaloes'' \citep{ghigna98,tormen04,gao04,delucia04,vdbtg}.
In what follows we will discuss an analytical approach to derive the 
mass function of subhaloes. We will use the simplifying assumption that no 
tidal stripping nor merging events among substructures happen.  
In this approach the mass of each subhalo remains constant in time, 
and equals the original virial mass \citep{eket96} of the progenitor halo at the considered 
redshift. A similar study was carried out by \citet{sheth03},
who calculated the subhalo mass function using the creation rate of the progenitors of a 
present day dark matter halo; our approach is different: we derive the subhalo mass 
function from the entire population of progenitors (as shown by Eq. \ref{nofprog}), 
in order to allow a direct comparison with the $N$-Body results of \citet{detal05}.

\subsection{Conditional Mass Function}
Let us consider a halo with virial mass $M$ at some final redshift $z_0$.
According to the hierarchical picture of galaxy formation, going backward 
in time the halo will be splitted in smaller and smaller systems,
called ''progenitors''. 
Mass conservation tells us that the sum of all masses of progenitor haloes 
at any given redshift equals the mass of the halo at $z_0$. 
Let us define the conditional mass function $f(m,z|M,z_0)\mathrm{d}m$ 
as the fraction of mass belonging to haloes with mass between $m$ and 
$m + \mathrm{d}m$ at redshift $z$, which are progenitors of a halo of mass 
$M$ (a $M$-halo) at a later redshift $z_0$.

Assuming the spherical collapse model \citep{ps74}, we can express 
$m$ and $z$ as a function of the new variables $s$ and $\delta_{sc}$. The 
conditional mass function is independent 
on the power spectrum of density fluctuations and it is described 
as \citep{lc93}:

\begin{equation}
  f(s,\delta_{sc}|S,\delta_{0})\mathrm{d}s =
  \frac{\delta_{sc}-\delta_{0}}{\sqrt{2 \pi(s-S)}}\,
  \exp \Big{\{}-\dfrac{(\delta_{sc}-\delta_{0})^{2}}{2(s-S)} 
  \Big{\}} \frac{\mathrm{d}s}{s-S}\,,
  \label{cmfps}
\end{equation}
where $s=\sigma^{2}(m)$ is the square of the mass variance of a $m$-halo, 
and $\delta_{sc}$ is the spherical collapse overdensity at redshift $z$. 
$S$ and $\delta_{0}$ are the mass variance of an $M$-halo and the spherical
collapse overdensity at the present time, respectively.
To compute the mass variance we have chosen a power spectrum with 
primordial spectral index $n=1$, and a transfer function obtained 
from CMBFAST \citep{sz96} for a concordance $\mathrm{\Lambda CDM}$ 
universe ($\Omega_{m}$, $\Omega_{\Lambda}$, $h$ = 0.3, 0.7, 0.7) 
with $\sigma_{8}=0.772$, extended down to a mass $M = 10^{6} M_{\odot}/h$. 

We have integrated this power spectrum using a top-hat filter in real space. 
To obtain  the mass variance until the typical Jeans neutralino mass 
we linearly extrapolate 
the $\log(m)$-$s$ relation to $M = 10^{-6} M_{\odot}/h$.

Over the last ten years $N$-Body simulations have shown that the
collapse of dark matter haloes is actually not well described by an
isolated spherical model; the influence of sourrounding proto-haloes 
can be reproduced using an ellipsoidal model \citep{smt01,st02}.

In the excursion set approach, the progenitor mass function of a halo is
described by the conditional probability of first upcrossing 
distribution.
Such a probability is well fitted by a random walk in the plane 
($s$, $\delta$), starting from ($S$, $\delta_{0}$) \citep{betal91}.

In the spherical collapse model this barrier has a constant height, 
defined by the collapse redshift: $B_{sc}(s,\delta_{sc})=\delta_{sc}$. 
For the ellipsoidal collapse case the barrier height is not constant,
but depends on $s$ and on $\delta_{sc}$ as described by the following 
equation:
\begin{equation}
  B_{ec}(s,\delta_{sc}) = \sqrt{q} \delta_{sc} \Big{[} 1 + \beta \Big{(}
  \frac{s}{q \delta_{sc}^{2}} \Big{)}^{\gamma} \Big{]}\,.
\label{barrier}
\end{equation}
\citet{smt01} found $q=0.707$, $\beta=0.5$ and $\gamma=0.6$; 
the value of the last two parameters is motivated by an analysis 
of the collapse of homogeneous ellipsoids, whereas the value of 
$q$ comes from requiring that the predicted halo abundances match 
what is found in the simulations.

Considering the barrier described in Eq.\ref{barrier}, 
\citet{st02} found an approximate 
solution for the diffusion equation, expressed as follows:
\begin{eqnarray}
  f(s,\delta_{sc}|S,\delta_{0})\mathrm{d}s &=&
  \frac{|T(s,\delta_{sc}|S,\delta_{0})|}{\sqrt{2 \pi(s-S)}} \times \\ \nonumber
  & &\exp\Big{\{}-\dfrac{[B(s,\delta_{sc})-B(S,\delta_{0})]^{2}}{2(s-S)}\Big{\}}
  \frac{\mathrm{d}s}{s-S}\,, 
  \label{cmfst}
\end{eqnarray}
with $T(s|S)$:
\[
  T(s,\delta_{sc}|S,\delta_{0})= \sum_{n=0}^{5} \frac{(S-s)^{n}}{n!} 
  \frac{\partial^{n}[B(s,\delta_{sc})-B(S,\delta_{0})]}{\partial s^{n}}\,.
\]

\begin{figure}
  \centering
  \includegraphics[width=\hsize]{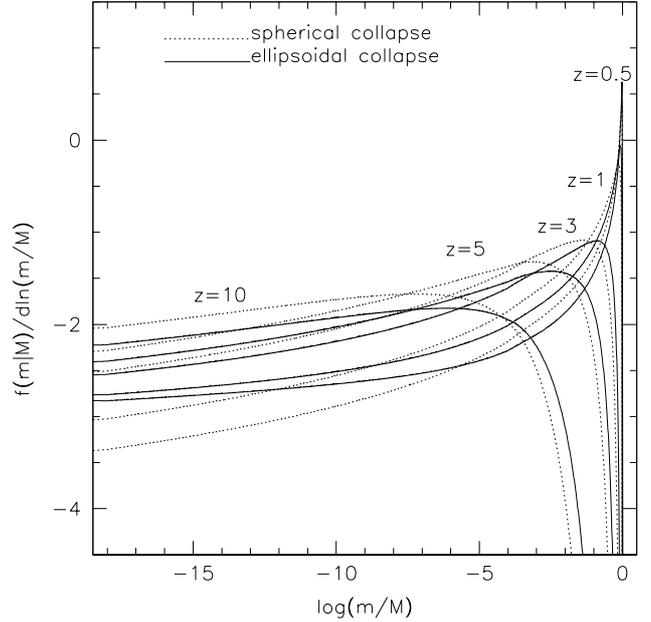}
  \caption{Ellipsoidal (solid) and spherical (dotted) conditional 
           mass function computed for a present-day dark matter 
           halo with mass $10^{12} M_{\odot}/h$ and for five different redshifts.\label{progmf}}
\end{figure}

In Figure \ref{progmf} we show the conditional mass function at five 
different redshifts for a halo with present-day mass 
$M = 10^{12} M_{\odot}/h$, both for the spherical (dotted curves) 
and ellipsoidal (solid) collapse  prediction. 
It can be observed that the halo 
is splitted in smaller and smaller progenitors at higher redshifts; 
discrepancies between the two 
models depend both on mass and on redshift.

Comparing the two prediction at fixed redshift, one can note that the spherical 
model predicts more progenitors at intermediate mass, and fewer at both very small
and very large masses, compared to the ellipsoidal model \citep{st02}. In other
words, the two predictions cross each others in two points, although these crossings
do not necessarily fall in the range of masses plotted in the Figure.

A direct consequence of this is that massive progenitors exist 
at higher redshifts in the ellipsoidal collapse, 
and the distribution of formation redshifts (defined as the 
earliest epoch when a halo assembles half of its final mass 
in one system) is consequently shifted to earlier epochs \citep{giocoli07}.

From $f(s,\delta_{sc}|S,\delta_{0})\mathrm{d}s$ we can write 
the total number of progenitors at any given redshift as:
\begin{equation}
  N(m,\delta_{sc}|M,\delta_{0}) \mathrm{d}m =
\frac{M(S)}{m(s)}f(s,\delta_{sc}|S,\delta_{0})\mathrm{d}s\,. \label{nofprog}
\end{equation}
Considering a scale free power spectrum $P(k) \propto k^{n}$, the mass variance 
scales as $s(m) \propto m^{-(n+3)/3}$, and the  number of progenitors 
can be explicitely written in terms of $s$:
\begin{equation}
  N(m,\delta_{sc}|M,\delta_{0}) \mathrm{d}m= \Big{(}\frac{s}{S}\Big{)}^{(n+3)/3} 
  f(s,\delta_{sc}|S,\delta_{0})\mathrm{d}s\,.
  \label{condwn}
\end{equation}	

\subsection{Number of progenitors}

Interating Eq. \ref{nofprog} over mass we obtain the total number of 
progenitors in the given mass interval, as a function of redshifts:
\begin{equation}
  \mathrm{d}n(z,\Delta m) = \int_{m_{i}}^{m_{f}}
  N(m,\delta_{sc}|M,\delta_{0}) \mathrm{d}m
  = \mathbb{N}(z) \Big{|}^{m_f}_{m_i}
    \,, \label{upsilon}
\end{equation}
where $m_{i}$ and $m_{f}$ represent the bounds of the interval.
For a white-noise power spectrum (scale free with $n=0$) and 
a spherical collapse mass function, a primitive of this integral 
can be written as:
\begin{eqnarray} 
  \mathbb{N}(z) &=& \frac{1}{S \sqrt{2 \pi}}
  \Bigg\{\mathrm{e}^{-\dfrac{(\delta_{sc}-\delta_{0})^{2}}{2(s-S)}} \\ 
  & &\Big{[} 2 \sqrt{s-S}(\delta_{sc}-\delta_{0}) -
  \mathrm{e}^{\dfrac{(\delta_{sc}-\delta_{0})^{2}}{2(s-S)}} 
  \nonumber \\ \nonumber
  & &\sqrt{2 \pi} [S - (\delta_{sc}-\delta_{0})^{2}] \mathrm{erf}\Big{(}
  \frac{\delta_{sc} - \delta_{0}}{\sqrt{2(s-S)}}\Big{)}
  \Big{]}\Bigg\}\,.
\end{eqnarray}

\begin{figure}
  \centering
  \includegraphics[width=\hsize]{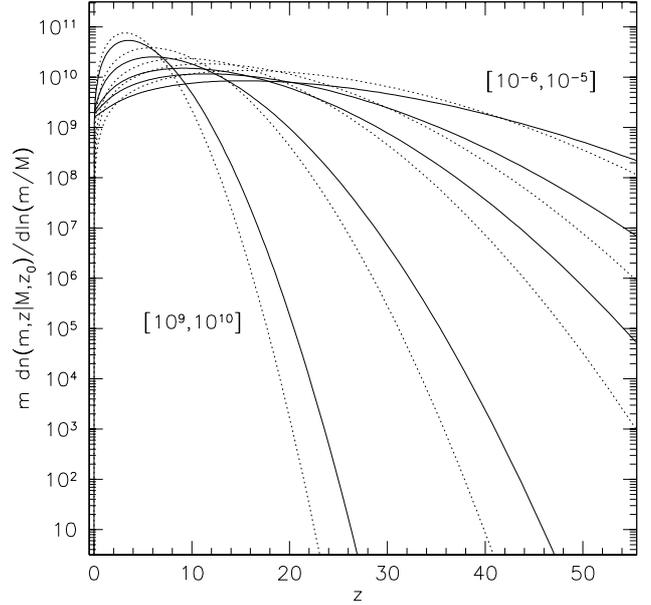}
  \caption{Total number of progenitors in a given mass bin, 
           as a function of redshifts, for a present day halo 
           with mass $M = 10^{12}\,M_{\odot}/h$. 
           For each mass bin we show the prediction for spherical
           (dotted lines) and ellipsoidal (solid) collapse models.
 \label{upsil}}
\end{figure}

In Figure \ref{upsil} we show the total number of progenitors in five
different mass decades, for a halo with mass $M=10^{12} M_{\odot}/h$ at
$z_{0}$, as a function of redshifts. We have assumed a concordance 
$\mathrm{\Lambda}$CDM power spectrum and have integrated Eq. \ref{nofprog}
numerically. 
The solid lines represent 
the prediction for the ellipsoidal collapse 
model while the solid ones refer to the ellipsoidal model. 
From top to bottom the curves represent the following mass 
bins: $[h \, 10^{-6},10^{-5}]$, $[10^{-1},1]$, 
$[10^{2},10^{3}]$, $[10^{6},10^{7}]$ and $[10^{9},10^{10}]$, all
but the first expressed in term of $M_{\odot}/h$.

It can be observed that the spherical collapse, for a
fixed mass bin, underpredicts the number of haloes at high redshifts
compared to the ellipsoidal model. We will see in the next sections
that if we consider the variable $\nu(z,m)=\delta_{sc}(z)/\sigma(m)$,
for any given mass this will result in the 
inequality $\nu_{ec}(m)>\nu_{sc}(m)$.

\section{Unevolved Subhaloes Mass Function from the merger tree 
         of a parent $M$-halo}
\label{sec:USMF}
The progenitors mass function, integrated over $\delta_{sc}$,
gives the total number of progenitors of mass between $m$ and
$m+\mathrm{d}m$ that a halo of final mass $M$ has had at all times:
\begin{equation}
  \frac{\mathrm{d}n(m)}{\mathrm{d}m} = \int_{\delta_{0}}^{\infty} \frac{M}{m} 
  f(s,\delta_{sc}|S,\delta_{0}) \mathrm{d} \delta_{sc}\,;
\end{equation}
in the case of the spherical collapse this integral results in:
\begin{equation}
  \frac{\mathrm{d}n(m)}{\mathrm{d} \ln(m)} = \frac{M}{\sqrt{2 \pi}} 
  \frac{|\mathrm{d} s / \mathrm{d}m |}{\sqrt{s-S}} \propto m^{-\alpha}\,,
\end{equation}
with $\alpha \approx 1 $ for a LCDM power spectrum. 
Since the same system may be a progenitor of the same final halo 
at more than one redshift, integrating the progenitor mass 
function overcounts the total number of progenitors. 
The result of this integration must then be properly 
re-normalized by imposing the constrain coming 
from \citep{dnat05} that roughly $10\%$ of the total Milky 
Way mass ($M=10^{12} M_{\odot}/h$) is in systems with mass 
ranging from $10^{7}$ to $10^{10}\,M_{\odot}/h$:
\begin{equation} 
\int_{10^{-5}}^{10^{-2}} \frac{m}{M} \mathrm{d} n = 0.1
\end{equation}
\begin{figure}
  \centering
  \includegraphics[width=\hsize]{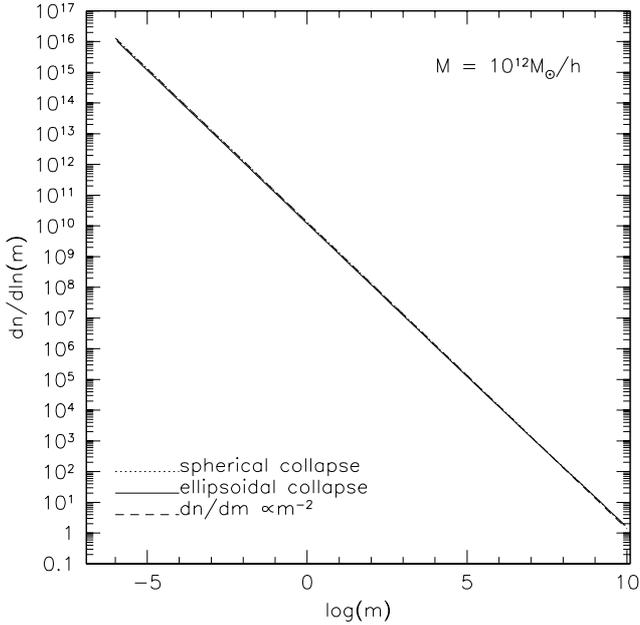} 
  \caption{Differential distribution of subhaloes in a
           $10^{12}\,M_{\odot}/h$ dark matter halo. The distribution has
           a slope approximatively equal to $1$ and has been normalized 
           considering that $10\%$ of the total mass is in subhaloes with 
           mass from $10^{7}$ to $10^{10}\,M_{\odot}/h$. \label{dn12}}
\end{figure}

In Figure \ref{dn12} we plot the differential mass distribution of
subhaloes in a $10^{12}\,M_{\odot}/h$ (Milky Way-like) dark matter
halo. The distribution has a 
power law behaviour approximately described by the relation:
\begin{equation}
  \frac{\mathrm{d}n(m)}{\mathrm{d}m} = A m^{-\gamma}\,,
\end{equation}
with $\gamma \approx 2$ for both the spherical and the ellipsoidal 
collapse model, respectively\footnote{A least-squares fit on the points
gives $\gamma_{\mathrm{sc}}=-1.9972 \pm 0.0001$ and 
$\gamma_{\mathrm{ec}}=-1.9937 \pm 0.0003$.}.
Once fixed the normalization factor, we find that the differential
distribution of the subhaloes is independent on the mass of the
progenitor halo, $M$, considering all the progenitors with mass 
from $10^{-6}\,M_{\odot}$ to $m/M=0.01$. 

\subsection{Progenitors $\sigma$-peak in the host halo}
\begin{figure}
  \centering
  \includegraphics[width=\hsize]{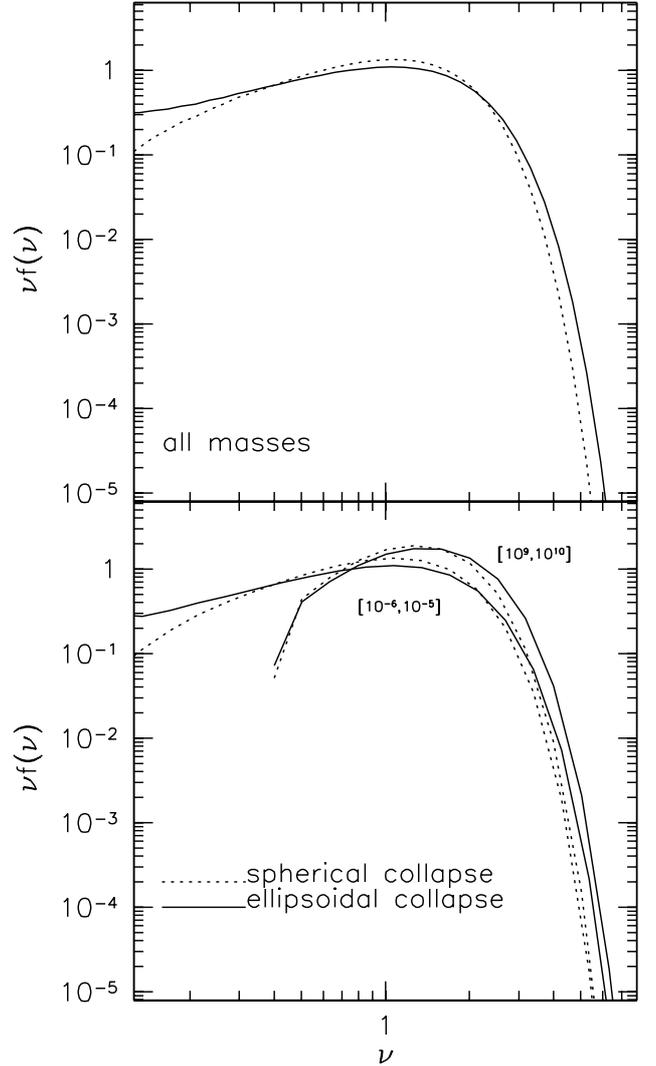}
  \caption{Progenitor mass function integrated over all redshifts. 
           In the top panel we show the distribution for all the 
           masses, while in the bottom panel we consider only
           progenitors in the first and last subhalo mass decades.\label{nuofnu}}
\end{figure}
Using high resolution $N$-Body simulations, \citet{detal05} studied
the spatial distribution - at $z=0$ - of matter belonging to high
redshift progenitors of a given system. They found that this
distribution mainly depends on the rareness of the density peak
corresponding to the progenitor, expressed in terms of $\nu =
\delta_{sc} / \sigma(M,z)$, and is largely independent on the
particular value of $z$ and $M$: matter from high $\nu$ progenitors
ends up at smaller distances from the center of the final system.

We can understand this in term of the revised secondary infall 
\citep{qz88,zar96}: the formation of haloes in $N$-Body
simulations preserves ranking of particle binding energy, that is,
particles in the cores of progenitor haloes will end up in the core of
the final system. Equally, particles from progenitors accreted at
earlier times, hence possessing more negative initial binding
energies, will likely have a more negative final energy, and so
be more centrally concentrated than average matter. 

At fixed redshift (hence at fixed $\delta_{sc}$), higher mass
progenitors have a larger $\nu$, are more self-bound than smaller mass
ones, and thus end up closer to the center of the final system.
Analogously, for a fixed progenitor mass, higher redshift progenitors
have a larger $\delta_{sc}$, hence a larger $\nu$; since at higher
redshift the universe is denser, they also are more self-bound than
lower redshift siblings, and so end up closer to the center of final
system.

In Figure \ref{nuofnu} we plot the subhalo mass function 
in terms of $\nu$. 
To compute the factor $\nu$ for each progenitor we integrated
the total number of progenitors in a given mass bin 
(Eq. \ref{upsilon}), at all redshifts. 
In the top panel we consider all the progenitors at all 
redshifts, with mass in the full range $h 10^{-6}$ to $10^{10}
\,M_{\odot}/h$; in the bottom panel we show the similar 
distribution only for the smallest and larger 
decade of the progenitors mass.

\section{$\gamma$-ray flux from Galactic substructures} 
\label{sec:flux}

\subsection{Modeling Galactic halo and substructures} 
\label{subsec:shape}
We model the distribution of DM in our Galaxy after \citet{detal05}.\\
For the smooth component of the Milky Way we use the best fit to the 
high resolution numerical experiments of \citet{dnat05}:
\begin{equation}
\rho_\chi(r) =
\frac{\rho_s} { \left( \frac{r}{r_s} \right)^\gamma \left [ 1 + 
\left( \frac{r}{r_s} \right)^\alpha \right]^{(\beta - \gamma)/\alpha}}
\label{density}
\end{equation}
with $(\gamma, \beta, \alpha) = (1.2, 3, 1)$. The scale radius 
$r_s$ and density $\rho_s$, are constrained by 
the virial properties of the halo. Following \citet{detal05} we adopt 
$r_s = 26 \kpc$, while $\rho_s$ has to be normalized to the virial
mass of the smooth DM halo.
We include a physical cutff $r_{cut}=10^{-8} \kpc$ which 
represents the distance at which the self-annihilation rate 
equals the dynamical time of spike formation. 

We shape the spatial distribution of subhaloes according to 
the fact that it traces the mass distribution of the parent 
halo from $r_{vir}$ down to a minimum radius $r_{min}(M)$ 
where tidal effects become important. We use 
Eq. \ref{density} together with the fact that the dependence from the initial 
conditions when the haloes accreted onto the present-day Milky 
Way halo is set through the parameter $\nu (M)$. 
We then use the parametrization obtained in \citet{detal05}:
$$
r_s \longrightarrow r_\nu = f_\nu r_s
$$
$$
f_\nu = \rm{exp}(\nu/2)
$$
\begin{equation}
\beta \longrightarrow \beta_\nu = 3 + 0.26 \nu^{1.6}
\label{nu_param}
\end{equation}
This parametrization reflects the fact that material accreted in 
areas with high density fluctuations is more concentrated toward 
the centre of the galaxy, and has a steeper outer slope. 
We also use the mass function derived in 
Sec.~\ref{sec:analyticalapproach} to model the number density 
of subhaloes per unit mass at a distance $r$ from the GC, for a 
given $\nu (M)$:   
\begin{equation}  
\rho_{sh}(M,r,\nu) = \frac{  A M^{-2} \theta (r - r_{min}(M))}
 { \left( \frac{r}{r_\nu (M)} \right)^\gamma \left [ 1 + 
 \left( \frac{r}{r_\nu (M)} \right)^\alpha \right]^{(\beta_\nu - \gamma)/\alpha}},  
\label{rho}  
\end{equation}  
in units of $\msun^{-1}  \kpc^{-3}$. The mass dependence in $r_\nu$ 
depends reflects the mass dependence of the virial parameter $r_s=r_{vir}/c_{vir}$.
The effect of tidal disruption is taken into account through 
the step function $\theta (r - r_{min}(M))$, where $r_{min}(M)$ is estimated 
following the Roche criterion.
A is a normalization factor obtained by imposing
that 10\% of the MW mass is distributed in subhaloes with masses in the range  
$10^{7}-10^{10} \msun$ \citep{dnat05}  as in Sec.\ref{sec:analyticalapproach}.

As a result about 50\% of the Milky Way mass is contained within
$\sim 2 \times 10^{16}$ subhaloes in the mass range 
[$10^{-6},10^{10}] \msun$.
The solar neighborhood density is $\sim 280 \pc^{-3}$, mainly
constituted by haloes with mass of $10^{-6} \msun$. 
The halo closest to the Earth is expected to be located $\sim 9.5 \times
10^{-2} \pc$ away. 

The remaining 50\% of the Milky Way mass 
is assumed to be smoothly distributed, and we use
this half mass value to normalize $\rho_s$ in Eq. \ref{density}. 

Few constraints exist on the density profile of each subhalo. 
Numerical simulations \citep{dnat05,dkm06,dkm07b} suggest they 
were formed with a NFW profile, which is described by
Eq. \ref{density} with  $(\gamma, \beta, \alpha) = (1, 3, 2)$. 
Even if subhaloes probably underwent tidal stripping and consequent 
mass loss after merging, their higher central density should prevent 
the inner regions from being affected.
\citet{PBB07} explored different possibilities for the 
concentration parameter $c_{vir} = r_{vir}/r_s$, where $r_{vir}$ 
is defined as the radius at which the mean halo 
density is $200$ times the critical density. 
Following their guidelines, we use two models for the 
concentration $c_{vir}$: we assume that the inner structure of 
subhaloes is either fixed at the time they merge onto the parent halo 
($z$-labeled model) or that it evolves with redshift until the present time
(0 model).
In model \B0 the NFW concentration is computed at $z=0$ according to \citet{Bullock:01} (hence the prefix $B$), and extrapolated to low masses.
In model \Bz, the values of $c_{vir} (M, z)$ are obtained from those at $z=0$ using the 
evolutionary relation $c_{vir}(M, z) = c_{vir}(M, z=0) / (1+z)$, where the
merging redshift $z$ is determined by the knowledge of the value of $\nu$ assigned to each progenitor.
 Therefore, 
subhaloes are much denser in model \Bz than in model \B0.

The values $c_{vir}$ thus found refer to progenitors formed from average density fluctuations 
($\nu = 1 \sigma$ peaks of the fluctuation density field). However, haloes with equal mass
at redshift $z_1$ may have assembled at different previous epochs; specifically, if we call
$z_i > z_1$ the redshift of mass assembly for progenitors observed at redshift $z_1$,
the amplitude of the initial density fluctuations producing the progenitors is an increasing
function of $z_i$. Therefore, their concentration $c(M,z)$ is also an increasing function of the
peak amplitude $\nu$. To account for this effect, we use the relation $c_{vir} (M, \nu)= \nu(M) c_{vir}(M,\nu=1)$, 
which has been tested against simulations by \cite{detal05}.

\subsection{Modeling the $\gamma$-ray flux from Dark Matter annihilation}
\label{subsec:flux}
We model the photon flux from neutralino annihilation 
in the population of galactic subhhaloes following \citet{PBB07}.
Given a direction of observation defined by the 
angle--of--view $\psi$ from the Galactic Center, and a
detector with angular resolution $\theta$, the $\gamma$-ray flux can be parametrized as:
\begin{equation}
\frac{d \Phi_\gamma}{dE_\gamma}(E_\gamma, \psi, \theta) =
\frac{d \Phi^{\rm PP}} {dE_\gamma}(E_\gamma) \times \Phi^{\rm
cosmo}(\psi, \theta)
\label{flussodef}
\end{equation}
The particle physics dependence in Eq. \ref{flussodef} is given by the
annihilation spectrum and DM properties and is embedded in the term:
\begin{equation}
\frac{d \Phi^{\rm PP}}{dE_\gamma}(E_\gamma) =  
  \frac{1}{4 \pi} \frac{\sigma_{\rm ann} v}{2 m^2_\chi} \cdot 
\sum_{f} \frac{d N^f_\gamma}{d E_\gamma} B_f. 
\label{flussosusy}
\end{equation}
$m_\chi$ is the DM particle mass, $\sigma_{\rm ann}v$ is the 
self--annihilation cross--section times the relative velocity 
of the two annihilating particles, and $d N^f_\gamma / dE_\gamma$ 
is the differential photon spectrum for a given final state $f$ with
branching ratio $B_f$, which 
we take from \citet{fps04}. 

The line--of--sight integral defined as:

$$ 
\Phi^{\rm cosmo}(\psi, \Delta \Omega) = \int_M d M \int_\nu d \nu \int \int_{\Delta \Omega}
d \theta d \phi \int_{\rm l.o.s}  d\lambda \int_c dc
$$
$$
[ \rho_{sh}(M,R(\rsun, \lambda,\psi, \theta, \phi), \nu) \times P(\nu(M)) \times P(c(M)) \times
$$
\begin{equation}
\times \Phi^{\rm cosmo}_{halo}(M,r(\lambda, \lambda ', \psi,\theta ', \phi '), \nu, c) \times J(x,y,z|\lambda,\theta, \phi) ]
\label{smoothphicosmo}
\end{equation}
accounts for the influence of cosmology in the flux computation.
$\Delta \Omega$ is the solid angle defined by the angular resolution of the instrument, $J(x,y,z|\lambda,\Delta \Omega)$ 
is the Jacobian determinant, $R = \sqrt{\lambda^2 + \rsun^2 -2
\lambda \rsun C}$, is the galactocentric distance and $r$ is the radial distance inside the single subhalo. $\rsun$ is the distance of the Sun from the
galactic center and $C=\cos(\theta) \cos(\psi)-\cos(\phi) \sin(\theta) \sin(\psi)$. $P(\nu(M))$ is the probability distribution function for the peak rarity $\nu(M)$ calculated using the extended Press-Schechter formalism.
$P(c(M))$ is the lognormal probability distribution for $c$ centered on
$c_{vir}(M)$ as it is computed in our models. While $P(\nu(M)$ is determined by the merging history of each subhalo, 
$P(c(M))$ describes the scatter in concentration for haloes of equal mass \citep{Bullock:01,neto}; therefore the two probabilities may be assumed independent. The single halo contribution to the total flux is given by
$$
\Phi^{\rm cosmo}_{halo}(M,r,\nu, c) = \int \int_{\Delta \Omega}  
d \phi ' d \theta '  \int_{\rm l.o.s} d\lambda '
$$
\begin{equation}
\left [ \frac{\rho_{\chi}^2 (M,r(\lambda, \lambda ',\psi,\theta ' \phi '), \nu,c)} {\lambda^{2}} J(x,y,z|\lambda ',\theta ' \phi ') \right] \, .
\label{singlehalophicosmo}
\end{equation}
This equation is also used to derive the contribution of the smooth component
of the MW itself.

Eq. \ref{smoothphicosmo} gives the average subhalo contribution 
to the Galactic annihilation flux within $\Delta \Omega$ along the 
direction $\psi$.

\begin{figure}
  \centering
  \includegraphics[width=\hsize]{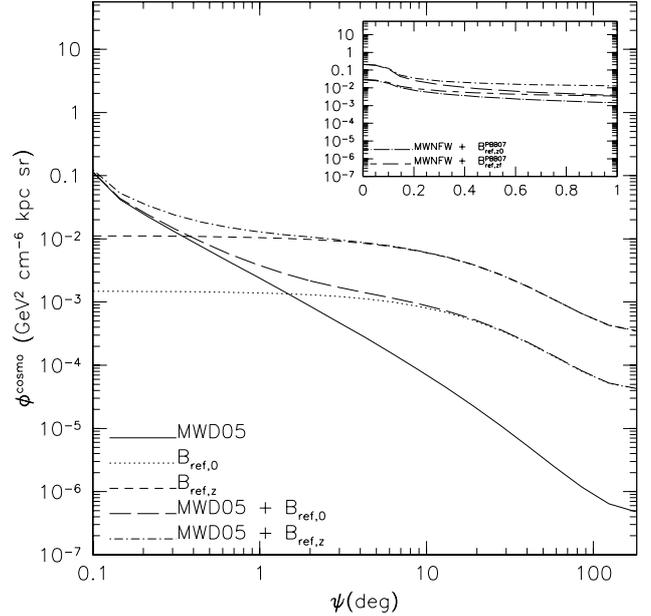}
  \caption{ Subhalo contribution to the $\gamma$-ray flux 
           for the two different models for the concentration parameters described in the text. MW smooth 
           and clumpy contributions are shown separately, together
           with their sum. In the small box, zoomed at small angles from the Galactic Center only the sum is shown, and it is compared with the values obtained in \citet{PBB07}. \label{fig5}}
\end{figure}

This contribution is shown in  Fig.~\ref{fig5}, together with the MW smooth halo component obtained with Eq.\ref{singlehalophicosmo}, for the two models considered in this analysis, for $\Delta \Omega = 10^{-5} \sr$, corresponding to an experimental angular resolution of $0.1^\circ$. The sum of the MW smooth and clumpy diffuse contributions is shown as well. We define this sum as our ''annihilation signal'', which will be multiplied by Eq.\ref{flussosusy} to obtain the predicted $\gamma$-ray diffuse flux from neutralino annihilation in our Galaxy.
In the small box we show a zoom at small angles of the annihilation signal and we superimpose the signal obtained in \citet{PBB07} for two similar models (we refer to their paper for the detailed explanation of models). Our models give a higher flux at the Galactic Center, where the signal is dominated by the MW smooth contribution. This is due to the different MW profile adopted. Yet, we find one order of magnitude of enhancement at the GC in the subdominant subhalo contribution as well, due to the presence of $P(\nu(M))$ in our determination of flux. Since more concentrated halos are closer to the GC in our approach, the enhancement is greater close to the GC: indeed, at the anticenter it goes 
down to a factor 2. 

We have used the $P(\nu(M))$ for the ellipsoidal collapse in Eq. \ref{smoothphicosmo}. We have checked that using the corresponding probability function for the spherical collapse does not change the result on $\Phi^{cosmo}$. This is due to the fact that the main difference between the two models resides at small values of $\nu$. A small $\nu$ gives low concentration parameter and its contribution to 
Eq. \ref{singlehalophicosmo} is then depressed with respect to that of a haloes with a higher $\nu$.

\subsection{Normalization to EGRET data} 
\label{subsec:norm}

\begin{figure}
  \centering
  \includegraphics[width=\hsize]{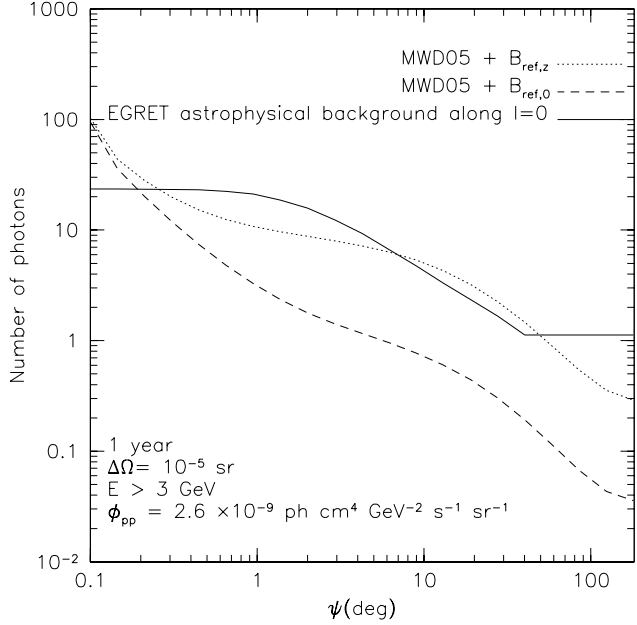}
  \caption{Number of photons above 3 GeV, in 1 year in a solid angle of $10^{-5} \sr$.
The annihilation signal models \B0 (dashed) and \Bz(dotted) are shown
together with the EGRET diffuse expected Galactic and extragalactic background (solid), as a function of the angle of view $\psi$ from the Galactic Center. \label{fig3}}
\end{figure}

In order to make predictions on detectability, we impose the best value of $\Phi^{PP}$ compatible with the available experimental limits.
As in \citet{PBB07}, we first assume the optimistic model where $m_\chi = 40 \ GeV$, $\sigma_{\rm ann}v = 3 \times 10^{-26} \ cm^3 \ s^{-1} $  and the branching ratio is $100\%$ in $b \bar b$. We then integrate Eq.\ref{flussosusy} above 3 GeV. 
This choice of parameters gives a value of  $\Phi^{PP}=2.6\times 10^{-9} \cm^4 \kpc^{-1} \GeV^{-2} \sec^{-1} \sr^{-1}$. \\
We then compute the expected number of photons above 3 GeV in 1 year for a solid angle of $10^{-5} \sr$ corresponding to the angular resolution of a GLAST-like satellite.
The result for the \B0 (dashed curve) and \Bz (dotted) models is shown in  Fig.~\ref{fig3}. \\

We compare the obtained number of events with the EGRET data for the diffuse Galactic component parametrized according to \citet{bck1}
\begin{equation}
\frac{d \phi^{\rm gal-\gamma}_{\rm diffuse}}{d\Omega dE}=
 N_0(l,b) \;10^{-6}\; E_{\gamma}^{-2.7} \frac{\gamma}{\cm^2 \sec \sr \GeV},
\label{dndegal}
\end{equation}
and with the diffuse extragalactic $\gamma$ emission, as
extrapolated from EGRET data at lower energies \citep{bck2}:
\begin{equation}
\frac{d \phi^{\rm extra-\gamma}_{\rm diffuse}}{d\Omega dE} 
= 1.38 \times 10^{-6} E^{-2.1} 
\frac{\gamma}{\cm^2 \sec \sr \GeV}. 
\label{gammas}
\end{equation}
The normalization factor $N_0$ in Eq. \ref{dndegal} depends only on the interstellar matter distribution.
The resulting number of photons above 3 GeV in 1 year for $\Delta \Omega = 10^{-5} \sr$, computed along l=0 where its value is minimum, is shown in  Fig.~\ref{fig3} (solid curve). \\
We find an excess of annihilation signal photons toward the Galactic centre in both models. Yet, the angular resolution of EGRET corresponding to $\Delta \Omega = 10^{-3} \sr$ does not allow to reconstruct a spiky source as it is ours. 
We have checked that, if we compute the number of annihilation signal photons toward $\psi=0$ smeared in a cone of view of $1^\circ$, it is below the number of EGRET detected photons for the same angular resolution. \\
Yet, the \Bz model exceeds the extragalactic diffuse measured background too, which is dominant above $\psi=40^\circ$. Since the extragalatic background 
is not due to any point source, we safely expect that it will scale with the solid angle.
The number of annihilation signal photons produced in the \Bz model should then be less or at most comparable with the number of measured background photons.
We make the optimistic assumption that the two numbers are comparable at $\psi=40^\circ$ where the discrepance is larger, and we thus fix $\Phi^{PP}_{B_{ref,z}}=2.0\times 10^{-9} \cm^4 \kpc^{-1} \GeV^{-2} \sec^{-1} \sr^{-1}$ for the \Bz model, correctly normalized to EGRET data, while we keep $\Phi^{PP}_{B_{ref,0}}=2.6\times 10^{-9} \cm^4 \kpc^{-1} \GeV^{-2} \sec^{-1} \sr^{-1}$ for the \B0 model.

\section{Prospects for detection}
\label{sec:detection}

In this section we study the sensitivity of a GLAST-like apparatus for 1 year of effective data taking. \\
We define the experimental sensitivity $\sigma$ as the ratio of the number $n_\gamma$ of annihilation signal photons and the
fluctuation of background events $n_{\rm bkg}$:
\begin{eqnarray}
\sigma &\equiv& \frac{n_{\gamma}}{\sqrt{n_{\rm bkg}}}\\ &=&
\sqrt{T_\delta} \epsilon_{\Delta \Omega}
\frac{\int A^{\rm eff}_\gamma (E,\theta_i) [d\phi^{\rm signal}_\gamma/dE
d\Omega] dE d\Omega}{\sqrt{ \int \sum_{\rm bkg} A^{\rm eff}_{\rm
bkg}(E,\theta_i) [d\phi_{bkg}/dEd\Omega] dE d\Omega}} \nonumber
\label{sensitivity}
\end{eqnarray}
where $T_\delta = 1$ year is the effective observation time and
$\phi_{bkg}$ is the background flux given by Eqs.\ref{dndegal} and \ref{gammas}, computed along l=0, that we assume to be composed by astrophysical photons only.
The quantity $\epsilon_{\Delta \Omega}$ is the fraction of signal
events within the optimal solid angle $\Delta \Omega$ corresponding to
the angular resolution of the instrument and it is optimistically set to 1.
$A^{\rm eff}$ is the effective detection area 
defined as the detection efficiency times the geometrical
detection area.
We use $A^{\rm eff} = 10^4 \cm^2$, independent
from the energy $E$ and the incidence angle $\theta_i$. 
Finally we assume an angular resolution of $0.1^{\circ}$ and an energy threshold of 3 GeV. \\

\begin{figure}
  \centering
  \includegraphics[width=\hsize]{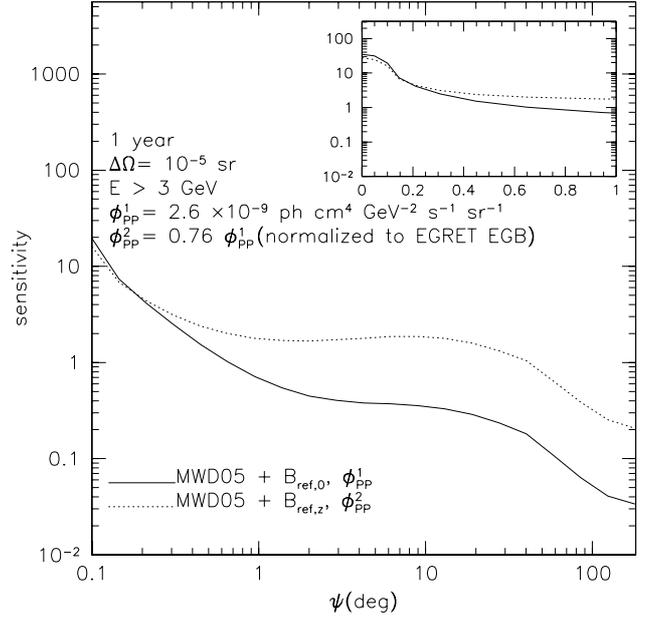}
  \caption{Sensitivity curves for a 
           GLAST-like experiment, for the \B0 (solid) and the \Bz (dotted) models described in the text.  A zoom at small angles is provided
           in the superimposed frame.\label{fig6}}
\end{figure}

The resulting sensitivity curves as a function of the angle of view $\psi$ are shown in Fig.\ref{fig6} for the \B0 (solid curve) and \Bz (dotted) annihilation signal models. 
In the small box a zoom at GC is shown.
An almost 2 $\sigma$ around $10^\circ$ is found for the \Bz model. The same model would be detected at about 30 $\sigma$ at the Galactic Centre.
As far as the \B0 model is concerned, it would show up with $\sim 40 \sigma$ effect toward the GC, that would rapidly fall down 1 $\sigma$ after $0.5^\circ$. A 5 $\sigma$ detection at the Galactic Center would be possible for both models with a value of $\Phi^{PP}$ even 6 times lower. 
In case of a striking excess detection along the GC, a milder excess a larger angles could be a hint for the discrimination about the models, though no discovery could be claimed.

\citet{PBB07} studied the detectability of resolved haloes which would shine above the Galactic foreground, finding in their best case scenario that only a tenth of large mass haloes would be detected, with a mass slope of -2 for the halo mass function.

Repeating their analysis is beyond the goal of this paper. Yet we note that the effect of including the $P(\nu(M))$ factor in Eq. \ref{smoothphicosmo} with respect to the concentration models in \citet{PBB07} leads to an enhancement of the Galactic foreground. We thus expect that including $P(\nu (M))$ will be compensated by
the increased foreground and 
we don't expect a dramatic change in the number of detectable haloes. \\
As a further test, we have computed the sensitivity of a GLAST-like experiment
for a \Bz halo once $\Phi_{PP}$ has been normalized to the EGRET data.
We chose the closer $M=10^{-6} \msun$ halo, located at $9.5 \times 10^{-2} \pc$ from the sun. We chose $\nu =2.4$ given from the probability of finding 1 halo with such a value in a 1 $\pc^3$ sphere around the sun. We conservatively considered only the astrophysical background in Eq. \ref{sensitivity}, while the annihilation signal foreground should be considered too. Even in these very optimistic hypothesis, we found that the source would produce a  5 $\sigma$ effect only furtherly multiplying by a factor of 4 the concentration parameter. This could be achieved using the lognormal probability $P(c(M))$ but with a ridicolously small probability. \\
We conclude that the effect of introducing the $P(\nu (M))$ can only be observed in a global enhancement of the diffuse Galactic annihilation foreground.

\section{Conclusions}
\label{sec:conclusions}

In this paper we have, for the first time, derived an analytical description 
of the mass function and distribution of rareness of density peaks in the 
subhalo population of our Galaxy, applying the extended Press \& Schecter formalism.
To make the calculation possible, tidal interactions and close encounters between subhaloes 
have been neglected. Very small (micro solar mass) subhaloes are extremely concentrated, 
therefore, at least for them, our approximation is a reasonable one.
 
The obtained results are valid over the whole range of subhalo masses 
$[10^{-6},10^{10}] \msun$ and thus confirm and extend the results of the 
N-body simulations, whose resolution is still far too low in order to simulate
coherently this mass range.

Making use of the results of \citet{detal05} on the distribution of different 
$\sigma$-peak material inside our Galaxy, we have been able to shape and model 
the total expected annihilation $\gamma$-ray foreground, statistically taking 
into account the merging history of each progenitor.

We have used the best case particle physics scenario to derive predictions
for the detectability of such a signal with a GLAST-like experiment.
We have shown how both the merging history and the intrinsic 
properties of the halo formation can contribute to an enhancement
of the expected flux, by arising the inner concentration of subhalos.
Yet the real concentration of the single subhalo today remains 
an open question. We use two models which result in very different inner densities inside 
the haloes. 
In the first model we assume that the inner shells of the subhaloes remain 
frozen at the moment they enter the parent halo and thus compute the 
concentration parameter at the merging epoch, as it is derived 
in our calculations. Alternatively we assume that the subhaloes continue to evolve 
with redshift, and thus compute the halo properties today.
We use the \citet{Bullock:01} model for the concentration parameter at $z=0$, 
extrapolated at low masses. We refer to \citet{PBB07} for the effect of using
different models.

Our results on detectability show that a detection would be
possible and impressive toward the GC for both models. 
This detection would be mainly due to the spike in the MW halo at the GC.
Unfortunately, a reliable modeling of the 
astrophysical background coming from the GC and of the effect of 
the central Super Massive Black Hole on the inner DM density profile 
are still poorly known.

A 2-$\sigma$ effect would show up as well, around $\sim 10^\circ$ from the GC, 
only for the \Bz model. Though no discovery could be claimed for,
this could be a significant hint for the existence of such a population of
subhaloes, and it would be propulsive for successive studies with upcoming
experimental technologies.

A final note on the metodology. In the present work we derived the final subhalo 
mass function starting from all progenitor haloes at any redshift. We did so
in order to directly compare our analytical results to the results obtained
by \citet{detal05} using $N$-Body simulations.  However, the subhalo population should
indeed be derived starting from the population of ''satellite haloes'' directly accreted
by the proto-halo (also called main progenitor) at all previous times \citep{torm97}, 
since only a fraction of progenitors at redshift $z$ merge directly with the main halo 
progenitor. Unfortunately, the mass function of satellite haloes cannot be obtained
analytically: it requires Monte Carlo simulations of the merging history tree of halo
formation \citep{sk99,vdb02,vdbtg}. We are currently working on this
issue (Giocoli et al, in prep.), and it will be interesting to compare
the results obtained using the two methods.

\bibliographystyle{mn2e}

\label{lastpage}
\end{document}